\documentclass[aps,prb,twocolumn,groupedaddress,floatfix,showpacs]{revtex4}
\usepackage{graphics}% Include figure files
\usepackage{epsfig}
\usepackage[latin1]{inputenc}
\usepackage{subfigure}
\usepackage{dcolumn}% Align table columns on decimal point
\usepackage{bm}% bold math
\usepackage{overpic} %insert latex on top of a picture

\begin{document}

\title{An experimental study of charge distribution in crystalline and amorphous Si nanoclusters in thin silica films}% Force line breaks with \\
\author{Annett Thøgersen}
\author{Spyros Diplas}
\author{Jeyanthinath Mayandi}
\author{Terje Finstad}

\affiliation{Department of Physics and Centre for Materials Science and
  Nanotechnology, University of Oslo, P.O.Box 1048 Blindern, N-0316 Oslo, Norway}

\author{Arne Olsen}
\affiliation{Department of Physics, University of Oslo, P.O.Box 1048 Blindern, N-0316 Oslo, Norway}

\author{John F. Watts}
\affiliation{Surrey Materials Institute and School of Engineering, University of Surrey, Guildford, Surrey GU2 7XH, UK}

\author{Masanori Mitome}
\author{Yoshio Bando}

\affiliation{National Institute of Material Science, Namiki 1-1, Tsukuba, Ibaraki, 305-0044 Japan}

\date{\today}

\begin{abstract}

\noindent Crystalline and amorphous nanoparticles of silicon in thin silica layers were examined by transmission electron microscopy (TEM), electron energy loss spectroscopy (EELS) and x-ray photoelectron spectroscopy (XPS). We used XPS data in the form of the Auger parameter to separate initial and final state contributions to the Si$_{2p}$ energy shift. The electrostatic charging and electron screening issues as well as initial state effects were also addressed. We show that the chemical shift in the nanocrystals is determined by initial state rather than final state effects, and that the electron screening of silicon core holes in nanocrystals dispersed in SiO$_2$ is inferior to that in pure bulk Si.

\end{abstract}



\maketitle

\section{Introduction}

\subsection{Si nanoparticles in SiO$_2$}

\noindent Drastic changes in materials properties and performance take place upon reduction of size and dimensionality of crystals and nanostructures. This has been the driving force in research and development of nanoscaled MOS (Metal-Oxide-Semiconductor) devices used for memory storage applications \cite{beltsios:mos, boer:mos}. Investigations on silicon nanocrystals (NCs) in silica matrices \cite{tiwari:memory} have been motivated by the possibility of replacing the original bulk-floating gate and applications \cite{pavesi:si} such as nanocrystal memory cells\cite{tiwari:si}, photon converters and optical amplifiers \cite{han:intro, kik:intro}. The main expectations include longer retention, lower gate voltage and lower power consumption \cite{chen:shift}. In addition, the discontinuity between the nanocrystals can prevent lateral charge loss and can also result in short writing times at lower voltages, as well as improved reliability \cite{ferdi:bias, lombardo:start}. Both injection and retention of electrons in these devices are very sensitive to the size, distribution, interfaces and electronic structure of the nanocrystals. Appropriate combination of these parameters may lead to dramatic improvements in device performances \cite{blauwe:start}.

\subsection{The Si$_{2p}$-shift in elemental silicon - earlier studies}

\noindent There is a significant number of studies on the Si/SiO2 system referring to planar or curved interfaces (particles in an oxide matrix), using XPS data. The majority of them was focused on the position and energy shift of the Si$^0_{2p}$ peak (were Si$^0$ is Si in elemental state) \cite{chen:shift, eickhoff:final, iwata:charge, chang:size, sun:charg1}. Studies of the planar SiO$_2$/Si(100) interface, attributed the shift either to final state relaxation effects\cite{eickhoff:final} or to enhanced differential charging by application of a negative bias to the substrate\cite{ferdi:bias}. Sun et al. \cite{chang:size} concluded that the shift is a result of an enlargement of the band gap due to surface imperfections at the Si nanocrystal-oxide interface. In a comprehensive paper from Iwata and Ishizaka\cite{iwata:charge} on planar SiO$_2$/Si interfaces, the shift was interpreted as a result of charging, that was dependent on x-ray intensity, time and sample thickness. They briefly mentioned that Auger parameters are independent of charging, without showing any implicit data. In a recent study, Dane et al. \cite{dane:auger} ascribed the shift in the Si$^0_{2p}$ binding energy of Si nanoclusters to relaxation energy differences measured by the final state Auger parameter.

\subsection{XPS chemical shift, the Auger parameter and chemical state diagrams}

\noindent X-ray Photoelectron Spectroscopy (XPS) is commonly used as a surface analysis technique to characterize chemical states of surfaces and interfaces. It is also frequently used for studying the electronic structure of materials. Interpretation of XPS spectra are often based on shifts of peak positions and Auger parameters \cite{evans:auger,elipe:ny}. The peak shift between two different chemical environments is known as chemical shift and is defined as the binding energy difference ($\Delta E_B$) between atoms bonded to different chemical species, e.g. elemental Si and SiO$_2$ (E$_B$ Si$^{4+}$ - E$_B$ Si$^0$). However, determination of chemical shifts depend on reliable measurements of XPS peak positions, which in turn are sensitive to energy referencing issues. Additional difficulties arise when measurements are performed on different samples or spectrometers and/or by using different experimental setups. Interpretations of the chemical shift are often complicated by differential charging when the sample is partially (semi)conducting and partially insulating. In addition, chemical shifts (in the form of binding or kinetic energy differences) contain both initial and final state contributions, as shown by equations \ref{eq:1} and \ref{eq:2} below\cite{evans:auger}. 

\begin{equation}
\label{eq:1}
\Delta E_B \sim \Delta V + \Delta \varphi - \Delta R
\end{equation}

 and 

\begin{equation}
\label{eq:2}
\Delta E_K \sim -\Delta V - \Delta \varphi + 3\Delta R 
\end{equation}

where ($\Delta E_B$) and ($\Delta E_K$) are the shifts in the photoelectron binding and Auger electron kinetic energy respectively. In the above formulas $\Delta$V reflects initial state changes in the atomic potential arising from changes in valence electron charge and/or Coloumb interactions at the emission site. The term $\Delta \varphi$ expresses changes in the work function and $\Delta$R refers to final state changes associated with relaxation/core hole screening energy.

The use of the Auger parameter ($\alpha$), as defined by equation \ref{eq:3} eliminates energy referencing problems \cite{wagner:auger}. 

\begin{equation}
\label{eq:3}
\alpha = E_B + E_K
\end{equation}

The combination of equations \ref{eq:1}, \ref{eq:2} and \ref{eq:3} leads to two different Auger parameter expressions reflecting either initial or final state effects; the initial\cite{evans:auger} ($\Delta \beta$) and final state\cite{Gaarenstrom:auger, weightman:auger} ($\Delta \alpha$) Auger parameter respectively: 

\begin{equation}
\label{eq:4}
\Delta \alpha = \Delta E_B + \Delta E_K = 2 \Delta R 
\end{equation}

and

\begin{equation}
\label{eq:5}
\Delta \beta = 3 \Delta E_B + \Delta E_K = 2( \Delta V + \Delta \varphi )
\end{equation}

The final state Auger parameter ($\alpha$) which is free of energy referencing problems, measures reliably the response of the system to the core hole electron screening \cite{walker:intro, matthew:intro}. The initial state Auger parameter is not completely independent of energy referencing due to the triple weighting of the binding energy ($E_B$) in its definition.

Very often chemical state or Wagner diagrams are employed to facilitate visualization and subsequent interpretation of the $\Delta E_B$ and $\Delta E_K$ values\cite{wagner:auger,moretti:start}. In the case of e.g. Si, such a diagram is constructed by plotting the binding energy of the Si$_{2p}$ peak against the kinetic energy of the Si$_{KLL}$ peak. The Auger kinetic energy is on the ordinate and the photoelectron binding energy is on the abscissa oriented in the negative direction. The Auger parameters are then expressed by the linear relationship E$_K$ (Auger peak) versus E$_B$ (photoemission peak) and lie on the straight lines with slope +1 (final state $\alpha$) and +3 (initial state$\beta$) \cite{moretti:start}. This means that all points lying on each line correspond to the same Auger parameter value. 

Due to the variety of results and explanations concerning the Si 2p shift (see section B above), we attempt a detailed study of the mechanisms resulting in the Si 2p shift using photoelectron spectroscopy data in the form of the Si$_{KLL}$-Si$_{2p}$ Auger parameter and chemical state (Wagner) plots, high resolution transmission electron microscopy (HRTEM) and electron energy loss spectroscopy (EELS). Electron Energy Loss Spectroscopy (EELS) is a useful tool in studying the electronic structure of nanostructural features of materials. Changes in the Si plasmon peak energy (due to valence electron vibrations) have been previously attributed to changes in quantum confinement and/or changes in the energy band gap \cite{mitome:plasmon}.

\section{Materials and methods}

\noindent The samples were produced by growing a ~3 nm layer of SiO$_2$ on a p-type silicon substrate by rapid thermal oxidation (RTO) at 1000$^o$C for 6 sec. Prior to growing the RTO layer the wafers were cleaned using a RCA (Radio Corporation of America) standard procedure for removing contaminants, followed by immersion in a 10 \% HF solution to remove the native oxide. A ~10 nm layer of silicon rich oxide was then sputtered from a SiO$_2$:Si composite target onto the RTO-SiO$_2$ and subsequently heat-treated in N$_2$ atmosphere at 1000-1100$^o$C for 30-60 min. Different percentages of sputtered material area coverage were achieved (Si:SiO$_2$ = 6, 8, 17,  42,  50, 60, 70\%), yielding different silicon supersaturation levels in the oxide. Table \ref{table:exp} gives information on the samples studied, their Si content and the applied heating duration and temperature. Cross-sectional TEM samples were prepared by ion-milling using a Gatan precision ion polishing system with a 5 kV gun voltage. 

\begin{table}[h!]
\caption{Silicon presentage, heat-treatment temperature and duration.}
\begin{tabular}{cccc}
\hline
\hline
Nr. in & area \% & Heating & Heat-treatment \\
 Wagner & Si & time & temperature \\
diagram (Fig. 6) & & (min) & ($^o$C)\\
\hline
1 & 100\footnotemark[1] & & \\
2 & 100\footnotemark[2] & 30 & 1000 \\
3 & 50 & 00 & 00 \\
4 & 60 & 00 & 00 \\
5 & 70 & 00 & 00 \\
6 & 28 & 30 & 1000 \\
7 & 42 & 30 & 1000 \\
8 & 42 & 60 & 1100 \\
9 & 50 & 30 & 1000 \\
10 & 50 & 120 & 1000 \\
11 & 60 & 30 & 1000 \\
12 & 70 & 30 & 1000 \\
13 & 70 & 60 & 1000 \\
14 & 70 & 60 & 1100 \\
\hline
\hline
\end{tabular}
\footnotetext[1]{Reference sample\cite{nist:mfp}}
\footnotetext[2]{Substrate}
\label{table:exp}
\end{table}

The Si nanoparticles (in crystalline and amorphous state) were observed by HRTEM and/or by energy filtered TEM- spectral imaging (EFTEM-SI) of the plasmon peak. HRTEM was performed with a 300 keV JEOL JEM-3100FEF TEM equipped with an Omega imaging filter. For EELS, a 200 keV JEOL 2010F TEM with a Gatan imaging filter and detector was used. Energy filtered images were acquired by filtering the plasmon peak of silicon (16.8 eV) with an energy slit of 2 eV. XPS was performed using a VG Scientific ESCALAB MkII fitted with a Thermo Electron Corporation Alpha 110 electron energy analyser, with non-monochromatic Al$_{Ka}$ radiation on plan view samples at a take-off angle of 45$^o$.  Survey scans and high resolution spectra were acquired using pass energies of 100 and 20 eV respectively. Non-monochromatic radiation was employed as it is the Bremsstrahlung component of the radiation from this type of source that is responsible for the excitation of the Si$_{KLL}$ transition \cite{castle:xps}. The inelastic mean free path of the Si$_{2p}$ electrons in SiO$_2$ is about 3.2 nm \cite{nist:mfp}. A take off angle of 45$^0$ results in a photoelectron escape depth of about 7 nm, which allowed us to study the silicon nanoclusters located 5 nm below the surface of the oxide. The spectra were peak fitted using Casa XPS \cite{casa:xps} after subtraction of a Shirley type background. The FWHM values used for fitting the various components of the Si$_{2p}$ peak Si, Si$_{2}$O, SiO, Si$_{2}$O$_{3}$ and SiO$_{2}$ were 1.4 eV, 1.6 eV, 1.9 eV, 1.9 eV and 2.0-2.3 eV respectively.

\section{Results}

\subsection{Crystalline and amorphous Si nanoclusters}

\begin{figure}
  \begin{center}
    \includegraphics[width=0.4\textwidth]{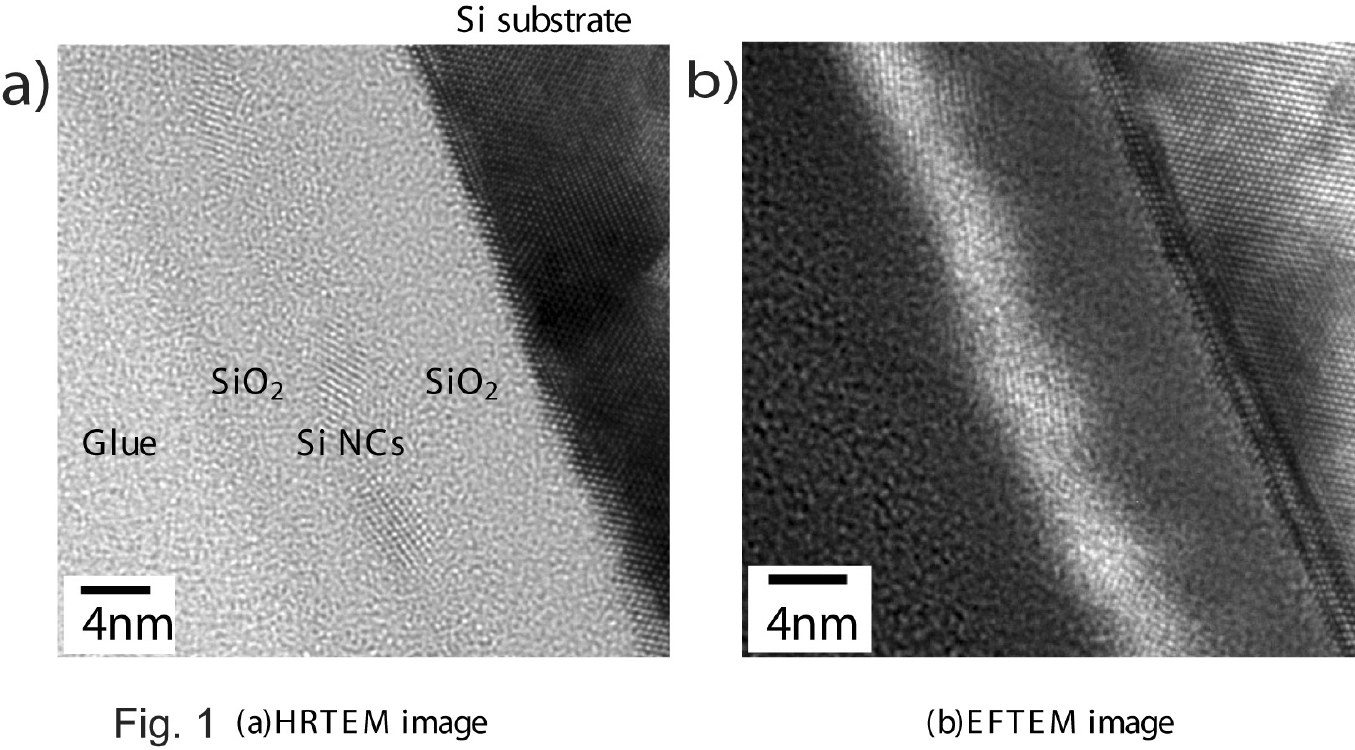}
    \caption{HRTEM and EFTEM-SI image of the sample with 70 area \% silicon, showing the nanocluster size and distribution. Images are representative for samples with 50, 60 and 70 area \% silicon.}
    \label{figure:1}
  \end{center}
\end{figure}

\noindent Samples with different area percentages of Si:SiO$_2$ have been systematically studied with HRTEM and energy filtered TEM (EFTEM) in order to relate the volume fraction, size and structure of nanocrystals to processing conditions. The results will be presented elsewhere\cite{Annett:senere}, However, the main findings are briefly summarized in the following two sentences in order to facilitate the discussion in the present work. HRTEM and EFTEM imaging showed presence of nanocrystals 3-8 nm in size in samples with a silicon fraction of 50 \% and higher (see also Figure \ref{figure:1}). Below 50 \% Si, amorphous nanoclusters of 3-6 nm were found.   

\subsection{Chemical states present in the samples}

\begin{figure}
  \begin{center}
    \includegraphics[width=0.4\textwidth]{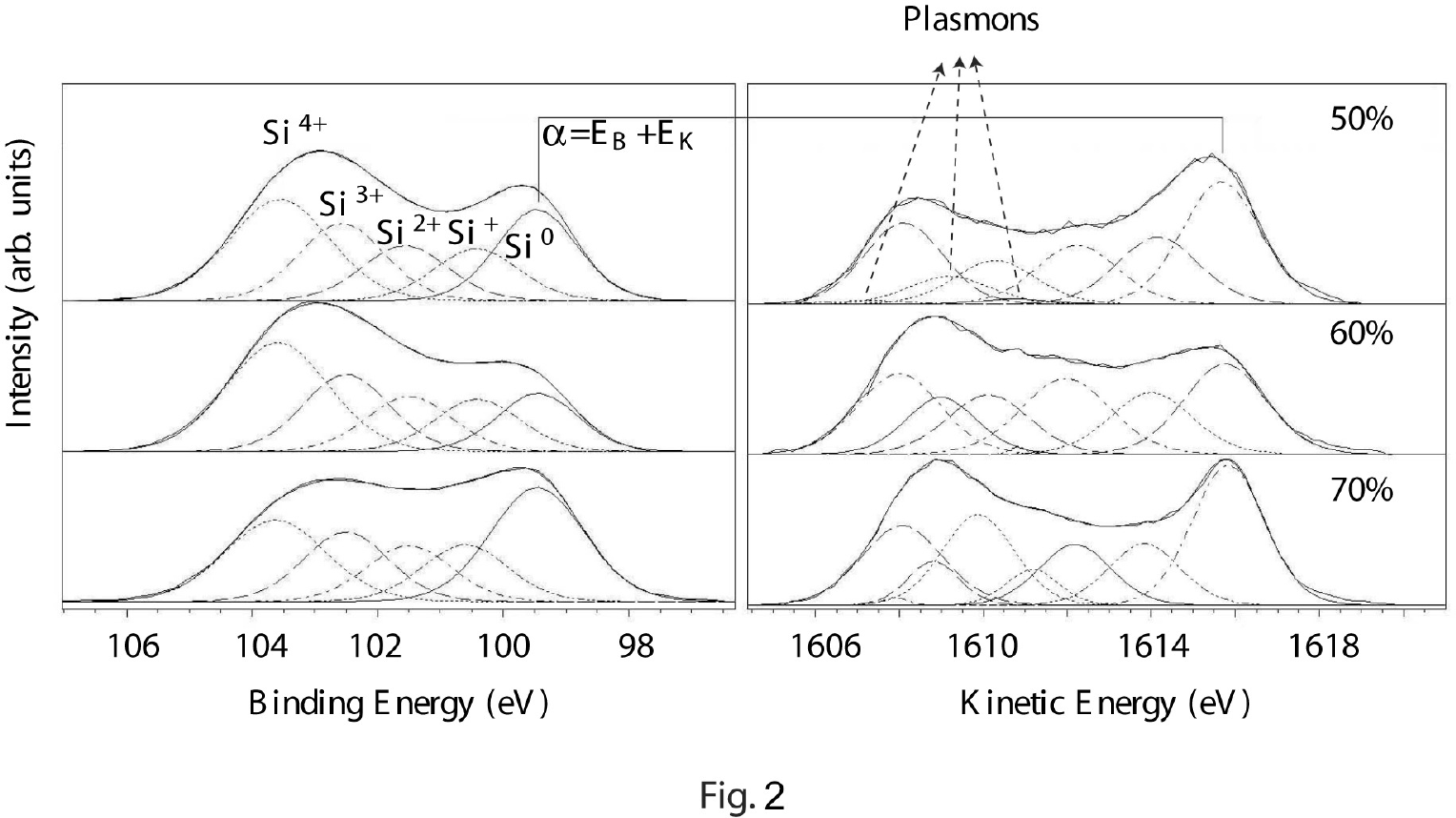}
    \caption{Si$_{2p}$ and Si$_{KLL}$ XPS spectra of the as grown samples with three different silicon concentrations, showing the chemical states present. The chemical shift is the same for all samples.}
    \label{figure:2}
  \end{center}
\end{figure}

\noindent XPS was utilized to identify the oxidation states of Si, detect the amount of elemental Si present in SiO$_2$, possibly as nanocrystals, and relate the shifts in core levels to Si concentration, nanocluster formation and size. Five oxidation states are reported to be present in Si/SiO$_2$ systems, corresponding to the five chemical states Si$^0$, Si$_{2}$O, SiO, Si$_{2}$O$_{3}$ and SiO$_{2}$ \cite{iwata:charge, 19:sio2, chen:shift}. Figure \ref{figure:2} shows the Si$_{2p}$ and Si$_{KLL}$ peaks of three as grown (A.G.) samples with different Si content showing the different chemical states present. The three extra peaks at 1610.9 eV, 1608.7 eV and 1607.8 eV in the Si$_{KLL}$ spectra are plasmon peaks of the Si$^0$, Si$^+$ and Si$^{2+}$ states and their use optimised peak fitting \cite{ivan:sikll, hirose:SiKLL}. The measured final state Auger parameters ($\alpha$) of Si$^0$ (E$_B$(Si$_{ref}^0$) = 99.5 eV) and Si$^{4+}$ (E$_B$(Si$_{ref}^{+4}$) = 103.6 eV) were 1715.4 eV and 1711.2 eV, respectively, in agreement with literature values 1715.7 \cite{smith:sio2} for Si$^{0}$ and 1711.7 eV \cite{wagner:sio2} for Si$^{4+}$. 

\begin{figure}
  \begin{center}
    \includegraphics[width=0.4\textwidth]{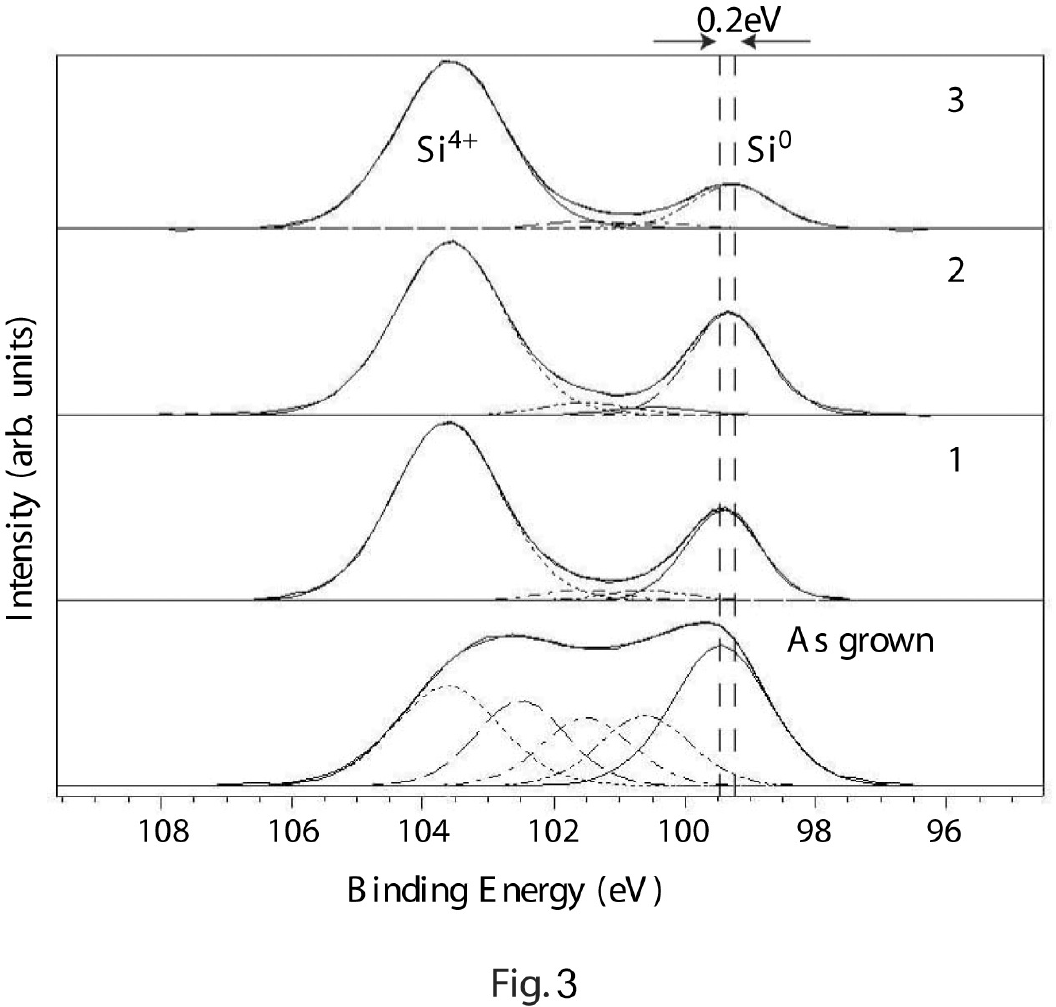}
    \caption{Si$_{2p}$ XPS spectra of the sample with 70 area percentage silicon, 1: 1000 $^o$C for 30 min, 2: 1000 $^o$C for 60 min, 3: 1100 $^o$C for 30 min. The Figure shows the chemical states and shifts and their variation upon heat treatment.}
    \label{figure:3}
  \end{center}
\end{figure}

During heat treatment of the as grown samples, the suboxides thermally decomposed to SiO$_2$ and Si$^0$ nanoclusters. It is clear from Figure \ref{figure:3} that annealing decreased the Si$^{3+}$-, Si$^{2+}$-, Si$^{+}$- and Si$^{0}$- content, and increased that of Si$^{4+}$. The reduction in intensity of the Si$^{0}$ peak may be attributed to oxidation due to residual oxygen in the nitrogen atmosphere \cite{chen:shift}. Comparison of the as grown sample with the sample annealed at 1100$^0$C for 30 min showed that annealing shifted the Si$^0_{2p}$ peak position to lower binding energy at 0.2 eV ($\pm$ 0.1). This decrease in binding energy between the as grown sample and the heat treated sample with a silicon fraction of 70 \% is also observed in samples with 50 and 60 \% silicon.

\subsection{Chemical shifts and nanoparticle size}

\noindent Table \ref{table:all} shows that the annealed sample with the highest Si fraction (70\%) contained 2.8 $\pm 0.1$ at. \% of suboxide, whilst the annealed sample with the lowest Si content (28 \%) had the highest amount of suboxides 5.3 $\pm 0.1$ at. \%. Therefore, it seems that the amount of suboxide increases with decreasing Si content and nanocluster size. This is probably an indication of the suboxide being located at the Si nanocluster/silica interface, created by Si ions not fully precipitated into the Si nanoclusters \cite{siox:1, siox:2, Daldosso:siox}.

\begin{table}[h!]
\begin{center}
\caption{Different parameters in crystalline nanoclusters (c-NC), amorphous nanoclusters (a-NC), as grown samples (A.G.), amorphous silicon and silicon substrate (Sub.) annealed at 1000$^o$C for 30 min. Asterisk denotes heat treatment for 2 hours.}
\begin{tabular}{cccccccccc}
\hline
\hline
area \% &  & & Mean & & Differential & & Chemical & & SiOx\\
Si & & & diameter & & Charging & & shift & &  \\ 
 & & & (nm) & & (eV)\footnotemark[1] & & (eV)\footnotemark[2] & & (atomic\%)\footnotemark[3]\\
\hline
28 & a-NC & & $<2$ & & 1.2 & & 5.2 & & 5.3 \\
42 & a-NC & & 4-6 & & 0.8 & & 4.9 & & 4.5 \\
50 & c-NC & & 4 & & 0.2 & & 4.4 & & 2.7 \\
60 & c-NC & & 4 & & 0.1 & & 4.3 & & 2.0 \\
50$^{*}$ & c-NC & & 5 & & 0 & & 4.2 & & 2.4 \\
70 & c-NC & & 4-8 & & 0.1 & & 4.2 & & 2.8 \\
\hline
ref\cite{ref:cs} & & & & & 0 & & 3.9 & & 0 \\
Sub. & & & & & 0.5 & & 4.6 & & 0 \\
a-Si\footnotemark[4] & & &  & & -0.4 & & 3.8 & & 0 \\
50 & A.G & & & & -0.9 & & 4.1 & & 20.7 \\
60 & A.G. & & & & -0.9 & & 4.1 & & 26.1 \\
70 & A.G. & & & & -1.1 & & 4.1 & & 22.4 \\
\hline
\hline
\end{tabular}
\footnotetext[1]{Differential charging:((E$_B$(Si$^{4+}$) - E$_B$(Si$^{4+}_{ref}$)) - (E$_B$(Si$^0$) - E$_B$(Si$^0_{ref}$)))}
\footnotetext[2]{Chemical shift:(E$_B$(Si$^{4+}$)-E$_B$(Si$^0$))}
\footnotetext[3]{Sum of suboxides}
\footnotetext[4]{Amorphous, hydrogenated silicon.}
\label{table:all}
\end{center}
\end{table}

\begin{figure}
  \begin{center}
    \includegraphics[width=0.4\textwidth]{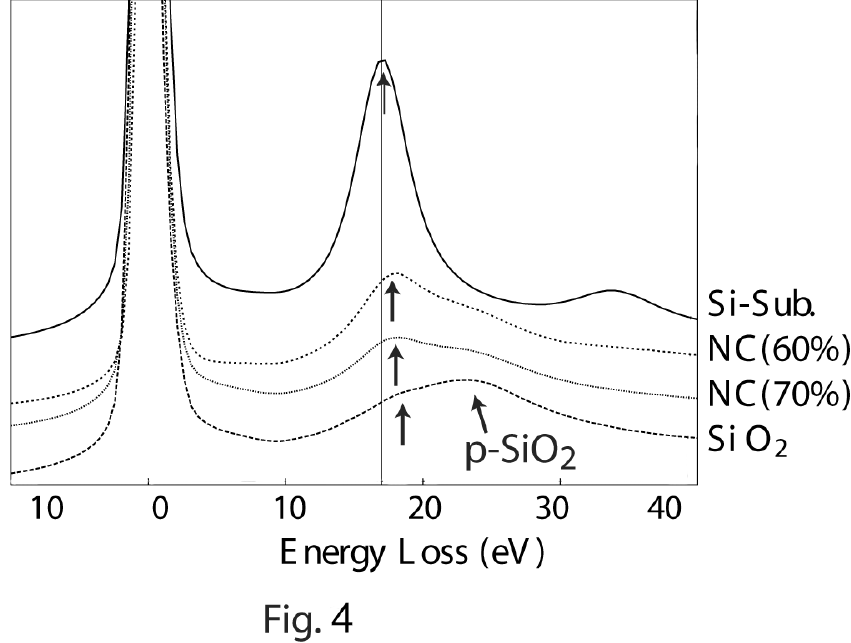}
    \caption{Si$_{2p}$ XPS spectra of the different samples annealed at 1000$^o$C for 30 min, showing differences in chemical shifts between the different samples. SiO$_2$ peaks were normalized for approximately the same binding energy and intensity. Asterisk denotes heat treatment for 2 hours. The as grown samples and the heat treated sample with a Si fraction of 28 \% have the smallest and largest shifts respectively.}
    \label{figure:4}
  \end{center}
\end{figure}

The Si 2p spectra in Figure \ref{figure:4} (normalized for same Si$^{4+}$ position) show differences in the chemical shift (Si$^{4+}$ - Si$^0$) between samples with different fraction of Si (see also Table \ref{table:all}). As grown samples (see also Figure \ref{figure:2}) and annealed samples with a Si fraction of 50 \% and above showed chemical shifts of 4.1 and 4.3 $\pm$ 0.1 eV, respectively. Samples with a lower fraction of Si contained amorphous Si clusters \cite{Annett:senere} and showed an increased chemical shift, whilst the measured shift for bulk Si (Si substrate revealed after 2-3 min of Ar etching) was 4.6 eV. These results illustrate a dependence of the shift on whether or not the samples contain amorphous or crystalline nanoclusters and also an increase of the shift with decreasing amorphous nanoclusters size.

\subsection{Initial state effects}

\noindent  Binding energy shifts between two different chemical environments contain initial state contributions due to the dependence of the potential on the local environment. They are also influenced by final state variations expressed as the relaxation energy change arising from the response of the local atomic electronic structure to the screening of the core hole. In order to address initial state variations we used the initial state Auger parameter and the Si plasmon peak energies and these values were compared to the measured differential charging.

The difference in electrostatic charging between silicon oxide and silicon nanocrystals/clusters (differential charging) is defined as\cite{chang:size} 

\begin{equation}
\label{eq:6}
(E_B(Si^{4+})-E_B(Si^{4+}_{ref}))-(E_B(Si^0)-E_B(Si^0_{ref}))  
\end{equation}

where E$_B$(Si$^{4+}$) is the Si$_{2p}$ binding energy for SiO$_2$, E$_B$(Si$^0$) is the Si$_{2p}$ binding energy for pure Si, and E$_B$(Si$^{4+}_{ref}$) and E$_B$(Si$^0_{ref}$) are the corresponding reference values \cite{anwar:nist}. Differences in Si$_{2p}$ peak positions without corrections for electrostatic charging together with reference values for Si$^{4+}_{2p}$  and Si$^0_{2p}$ are shown in Table \ref{table:all} \cite{xray:ref}. The samples containing nanocrystals (Si fraction of 50 \% and above), showed the smallest differential charging (0.1 eV), whilst the sample containing amorphous nanoclusters (with the lowest Si concentration), exhibited the highest one (1.2 eV). Taking into account the results showed in secion C, the differential charging increases with increasing chemical shift. Samples containing nanocrystals show the same chemical shift and the same differential charging irrespective of the nanocrystal size whilst for the samples with amorphous nanoclusters (28, 42\% Si) the differential charging increases with decreasing nanocluster size.  

\begin{figure}
  \begin{center}
    \includegraphics[width=0.4\textwidth]{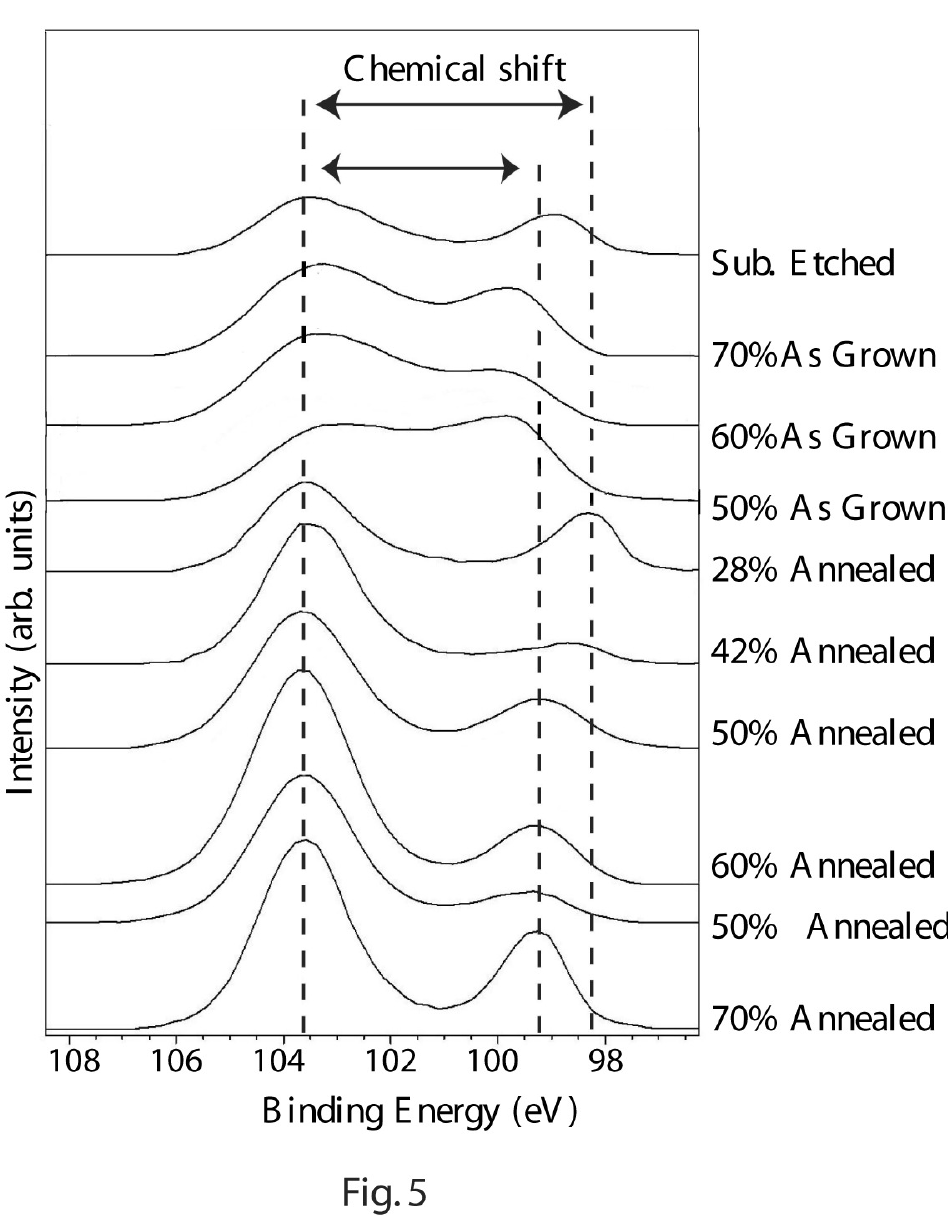}
    \caption{The difference in plasmon peak energy of pure silicon between the substrate, nanocrystals and the SiO$_2$ of the sample with a Si fraction of 60 and 70 \% (shown by arrows). The increase in intensity of the plasmon tail at higher energies is due to presence of SiO$_2$ (p-SiO$_2$).}
    \label{figure:5}
  \end{center}
\end{figure}

EELS-low loss spectra from the Si substrate, the SiO$_2$ and the Si nanocrystals in samples with 60 and 70 \% Si are shown in Figure \ref{figure:5}. The maximum in the plasmon energy peak for pure bulk Si and the Si nanocrystals in the 70 and 60 \% samples were 17.2 eV, 18.4 eV and 18.8 eV respectively. A 1.2-1.6 eV difference in plasmon energy between pure Si (substrate) and the Si nanocrystals was therefore observed. Previous studies have attributed the difference in plasmon peak energy between nanocrystals with different sizes to quantum confinement and/or changes in the energy band gap \cite{mitome:plasmon}. 

\begin{figure}
  \begin{center}
    \includegraphics[width=0.4\textwidth]{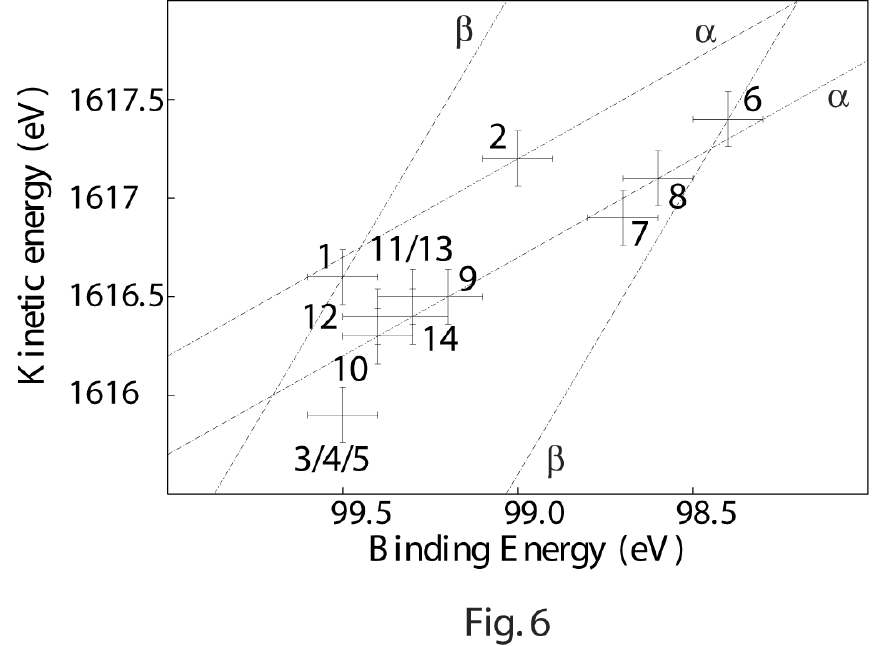}
    \caption{A Wagner diagram showing the binding energy of the Si$_{2p}$ peak plotted against the kinetic energy of the Si$_{KLL}$ peak. The Auger kinetic energy is on the ordinate and the photoelectron binding energy is on the abscissa oriented in the negative direction. The Auger parameters are the intercepts of the linear relationship E$_K$ (Auger) versus E$_B$ (photoemission) to be read directly on the straight line with slope +1 (final state) and +3 (initial state). All points lying on each line have the same Auger parameter.}
    \label{figure:6}
  \end{center}
\end{figure}

The initial state Auger parameter ($\beta$) was measured using equation \ref{eq:5} and the values are shown in table \ref{table:auger} and plotted in the Wagner diagram in Figure \ref{figure:6}, with the values lying on the straight lines with slope +3. The reference values\cite{xray:ref} used in these plots are Si$_{KLL}$(Si$^0$) = 1616.6 eV, Si$_{KLL}$(SiO$^{+4}$) = 1608.8 eV , Si$^{4+}_{2p}$ = 103.6 and Si$^0_{2p}$ = 99.5. The Wagner diagram (Figure \ref{figure:6}) shows a large deviation in $\beta$ between samples with different fractions of Si and heat treatment. Samples containing small, amorphous nanoclusters (e.g. the sample with 28 \% Si) have the lowest value for $\beta$ (1912.6 eV), which results in a $\Delta \beta$ ($\Delta \beta$ = $\beta_{sample}$-$\beta_{ref}$) of -2.5 $\pm$ 0.4eV, as compared to pure bulk Si. As grown samples and samples with nanocrystals have the largest $\beta$ values, 1915.7 eV ($\Delta \beta$ is -0.7 $\pm$ 0.4 eV) and 1914.6 $\pm$ 0.3 eV ($\Delta \beta$ of -0.5 eV, for sample with 70 \% silicon) respectively . Changes in $\beta$ are a measure of the ground state chemistry, changes in environmental potential and charge transfer \cite{bagus:strain}. More specifically, negative $\Delta \beta$ value is associated with a negative shift of the potential ($\Delta$V) \cite{evans:auger, moretti:start}. In the above context, the decrease in initial state Auger parameter ($\beta$) of the amorphous nanoclusters indicates an increased accumulation of negative charge as compared to the crystalline nanoclusters.

\begin{table}[h!]
\caption{Values of the initial state Auger parameter ($\beta$) and the final state Auger parameter ($\alpha$) for samples with different fractions of Si.}
\begin{tabular}{cccccc}
\hline
\hline
Nr. in & area \% & $\alpha$ -Si & $\Delta \alpha$& $\beta$ -Si & $\Delta \beta$\\
 Wagner & Si & $\pm$ 0.1 eV & $\pm$ 0.1 eV & $\pm$ 0.3 eV & $\pm$ 0.4 eV \\
diagram & & & & & \\
\hline
1 & 100\footnotemark[1] & 1716.1 & 0 & 1915.1 & 0 \\
2 & 100\footnotemark[2] & 1716.2 & 0.1 & 1914.2 & -0.9 \\
3 & 50 & 1715.4 & -0.7 & 1914.4 & -0.7\\
4 & 60 & 1715.4 & -0.7 & 1914.4 & -0.7 \\
5 & 70 & 1715.4 & -0.1 & 1914.4 & -0.7 \\
6 & 28 & 1715.8 & -0.3 & 1912.6 & -2.5 \\
7 & 42 & 1715.6 & -0.5 & 1913.0 & -2.1 \\
8 & 42 & 1715.7 & -0.4 & 1912.9 & -2.2 \\
9 & 50 & 1715.7 & -0.4 & 1914.1 & -1.0 \\
10 & 50 & 1715.7 & -0.4 & 1914.5 & -0.6 \\
11 & 60 & 1715.8 & -0.3 & 1914.4 & -0.7 \\
12 & 70 & 1715.8 & -0.3 & 1914.6 & -0.5 \\
13 & 70 & 1715.8 & -0.3 & 1914.4 & -0.7 \\
14 & 70 & 1715.7 & -0.4 & 1914.3 & -0.8 \\
\hline
\hline
\end{tabular}
\footnotetext[1]{Reference sample\cite{nist:mfp}}
\footnotetext[2]{Substrate}
\label{table:auger}
\end{table}

\subsection{Final state effects}

Final state effects express properties of the short lived, highly excited core hole states and are related to screening efficiency \cite{snyder:screening}. The final state Auger parameter ($\alpha$) provides an estimate of the relaxation/screening energy in the presence of core holes, (see equation \ref{eq:4}). A high $\alpha$ value indicates higher relaxation energy or improved screening efficiency. Final state Auger parameter values are shown in Table \ref{table:auger} and plotted in the Wagner diagram shown in Figure \ref{figure:6} as lines with slope +1.

The variations in $\alpha$ between the different samples are smaller than in $\beta$. The only significant difference in $\alpha$ is between bulk Si (substrate and literature \cite{nist:mfp} value of pure Si) and the nanoclusters (Si$^0_{NC}$), as well as between bulk Si (Si$_{bulk}^0$) and the silicon in the non annealed samples (Si$^0_{AG}$). Bulk silicon has the highest $\alpha$ value (1716.2 $\pm$ 0.1 eV) in agreement with the literature value for pure silicon, whilst Si$_{AG}^0$ has the lowest Auger parameter (1715.4 $\pm$ 0.1 eV). The above results indicate that electronic screening of core holes increases in the order Si$^0_{AG}<$Si$^0_{NC}<$Si$^0_{bulk}$.

\section{Discussion}

\subsection{Initial state effects}

\noindent Contrary to $\alpha$, the Auger parameter $\beta$ is not completely independent of energy referencing. Therefore, $\beta$ can be potentially influenced by work function differences. The work function is defined as the work necessary to remove a Fermi-level (E$_F$) electron from the crystal to infinity \cite{weinert:wf, wigner:wf}. E$_F$ is the level at which the electrons and protons are balanced, halfway between the valence and conduction bands in intrinsic semiconductors, and it shifts either towards the conduction band, or towards the valence band depending on doping, n and p type respectively. 

Since experiments were performed using the same spectrometer, we assume that the effect of spectrometer work function on binding energy ($\Delta$E$_B$) should be negligible. Therefore it can only be differences in the material work function that may contribute to $\Delta$E$_B$. An increase in energy band gap E$_g$ could shift E$_F$ upwards towards the vacuum level, thus reducing the work function ($\varphi$). This would be detectable via an increase in the E$_K$ of the $Si_{2p}^0$ electrons, an equivalent reduction in E$_B$ and a subsequent increase in the chemical shift. An increase in valence charge could also lift up the E$_F$ and therefore reduce $\varphi$.

According to the suggested interpretation, plasmon energies and Auger parameter values provided by EELS and XPS respectively, indicate an increase in valence electron density and/or E$_g$ in the amorphous nanoclusters compared to bulk silicon. Either one or both of the two reasons (increased electron density, larger E$_g$) lift the E$_F$ and subsequently reduce $\varphi$ of the nanocrystals, influencing thus (lowering) the Si$_{2p}^0$ binding energy and increasing the chemical shift.

Pure silicon (Si$^0$) supersaturated in SiO$_2$ prior to annealing shows exactly the opposite behavior in Si$_{2p}^0$ peak shifts, chemical shift and differential charging than the amorphous nanoclusters (see Table \ref{table:all}). Also the initial state Auger parameter of Si$^0$ in SiO$_2$ prior to annealing exhibits only a small negative shif compared to the substrate (see Table \ref{table:auger}). This implies that the ground state (Si$^0$) of silicon in SiO$_2$ is ``more positively charged'' prior to annealing and becomes ``more negatively charged'' upon annealing during formation of amorphous and crystalline nanoclusters. The positive shift ($\Delta$V) of the atomic potential of Si$^0$ in the as grown samples (compared to Si$^0_{bulk}$) may be attributed to the need of Si to share electrons with the more electronegative surrounding oxygen atoms. When surrounded by the less electronegative Si atoms, (Si$^0$ atomic environment in nanocrystals), the electrons are located closer to the Si atom.

\subsection{Final state effects}

The increase in band gap would also lead to a reduction in the screening efficiency of  the conduction electrons and this will be detected as a decrease in $\alpha$. Small nanocrystals have an increased band gap compared to both bulk silicon and larger nanocrystals \cite{brus:gap, kanemitsu:gap, kayanuma:gap}, and this is reflected in $\alpha$ aquiring lower values \cite{moretti:egap}. Amorphous Si has a higher band gap (1.6-1.7 eV \cite{tanenbaum:asi, chu:asi}) than crystalline Si (1.2 eV). We recall that amorphous nanoclusters also have a larger difference in chemical shift, differential charging and initial state Auger parameter, compared to the nanocrystals and the substrate. The final state Auger parameter $\alpha$ shows only small differences between crystalline bulk Si and Si nanoclusters (both crystalline and amorphous), but there is no significant difference between the differently sized amorphous or crystalline nanoclusters, therefore it seems that final state effects influence the Si$_{2p}$ binding energy less than initial state effects (see Table \ref{table:auger}). The $\alpha$ for silicon in both the as grown and post annealed conditions is lower than that in bulk silicon or reference Si, implying a reduced core hole screening. It is known that screening in Si has a non local character \cite{weightman:auger2}. Therefore the delocalized nature of electron screening in combination with the non-conducting silicon oxide environment would be expected to influence the screening of core holes in the nanoclusters even in the presence of an increased valence electron density.

\section{Conclusions}

\noindent We used the Auger parameter to separate initial and final state effects in the Si$_{2p}^0$ shift of Si crystalline and amorphous nanoclusters dispersed in SiO$_2$. The Si$^0_{2p}$ position in the nanocrystals relative to Si$^{4+}_{2p}$ (chemical shift) is determined by initial state rather than final state effects. The negative charge on the Si$^0$ sites in nanoclusters and the positive on the Si$^0$ sites in the supersaturated (with Si) silica, as indicated by the initial state Auger parameter, dominate on both differential charging and chemical shifts. $\Delta \alpha$ shows that the electron screening of core holes in Si is superior when Si is clustered and not dispersed in SiO$_2$. The core hole screening of Si in the nanoclusters is inferior to that in bulk Si and this is presumably due to its non local character in Si.

\section{Acknowledgement}

\noindent Financial support by FUNMAT-UiO, the University of Oslo and Kristine Bonnevie's travelling scholarship is greatfully acknowledged. We are also grateful to the referee for his comments.

\end{document}